\documentclass{aastex}          
\usepackage{spr-astr-addons}    

\begin{document}
%
  \title{IDV Observations \& Study of the Quasar 0917+624}

\shorttitle{IDV Study of Quasar 0917+624}

\shortauthors{Xiang Liu et al.}

\author{Xiang Liu\altaffilmark{1,2}}
\and
\author{Quanwei Li\altaffilmark{1,3}}
\and
\author{Thomas P. Krichbaum\altaffilmark{4}}
\and
\author{Margo F. Aller\altaffilmark{5}}
\and
\author{Hugh D. Aller\altaffilmark{5}}

\email{liux@xao.ac.cn}

\altaffiltext{1}{Xinjiang Astronomical Observatory, Chinese
Academy of Sciences, 150 Science 1-Street, Urumqi 830011, PR
China}

\altaffiltext{2}{Key Laboratory of Radio Astronomy, Chinese
Academy of Sciences, Nanjing 210008, PR China}

\altaffiltext{3}{University of Chinese Academy of Sciences,
Beijing 100049, PR China}

\altaffiltext{4}{Max-Plank-Institut f\"ur Radioastronomie, Auf dem
H\"ugel 69, 53121 Bonn, Germany}

\altaffiltext{5}{Department of Astronomy, University of Michigan,
Ann Arbor, MI 48109-1042, USA}

\begin{abstract}
We carried out intra-day variability (IDV) observations from
August 2005 to January 2010 with the Urumqi 25m radio telescope
for a dozen IDV sources including the quasar 0917+624. This target
exhibited pronounced centimeter-band, intra-day variability during
the 1980s--1990s, but its strong IDV phase ceased in 2000. The
source showed no IDV in the majority of the Urumqi observing
sessions, although weak IDV activity was detected in some.
Multifrequency UMRAO data for 0917+624 show that the spectral
index is steeper during the weak and non-IDV phases than during
the strong IDV phase, supporting the idea that the size of the
scintillating component may be enlarged in the weak/non IDV
phases.
\end{abstract}


\section{Introduction}

It is reported in the literature that $\sim$25\% to  $\sim$50\% of
flat-spectrum radio sources \citep{qui92, lov08} and $\sim$60\% of
bright {\it Fermi} blazars \citep{liu11, liu12a} exhibit
radio-band intra-day variability (IDV). The quasar 0917+624
(OK630, z=1.453) was one of the most pronounced centimeter-band
IDV sources during the 1980s--1990s \citep{hee87}. However, its
strong IDV activity stopped from 2000 to 2001 \citep{fuh02}. The
data during the strong IDV phase before 2000 have been analyzed
assuming an ISS model of our galaxy, and the results are
consistent with an ISS origin for the IDV \citep{jau01, ric01}.
However, source-intrinsic variations cannot be completely ruled
out, because of the remarkable polarization variations detected in
the IDV of 0917+624 \citep{qui89, qian91}; the polarization
variations were explained with some success using an ISS model by
\citet{ric95} (where the source has more than one polarized
component). To investigate IDV in recent years, we carried out IDV
observations for
a dozen sources at Urumqi and present results for 0917+624 in this research note.\\

\section{Observations and results}

The quasar 0917+624 was observed at Urumqi from August 2005 to
January 2010 at 4.8 GHz, with approximately monthly observations
of $\sim$4 days per session. The observing strategy and detailed
calibration method is described in \citet{liu12b}. The relative
uncertainty of the calibrated total flux density variability is
around 0.5\%, in normal weather conditions.

In order to ascertain whether or not IDV is present or whether the
flux is constant during each IDV observing session for 0917+624 we
applied a $\chi^{2}$ test to each dataset. We adopt the criterium
that a dataset with a probability of $\leq$\,0.1\% to be constant
is considered to be variable; i.e., we set the confidence level of
99.9\% to define the IDV. From this analysis, we find no IDV in
the majority of the sessions and weak IDV in some sessions at a
confidence level of $\geq\,99.9$\,\%.
 \begin{figure}
   \includegraphics[width=10cm,height=7.5cm]{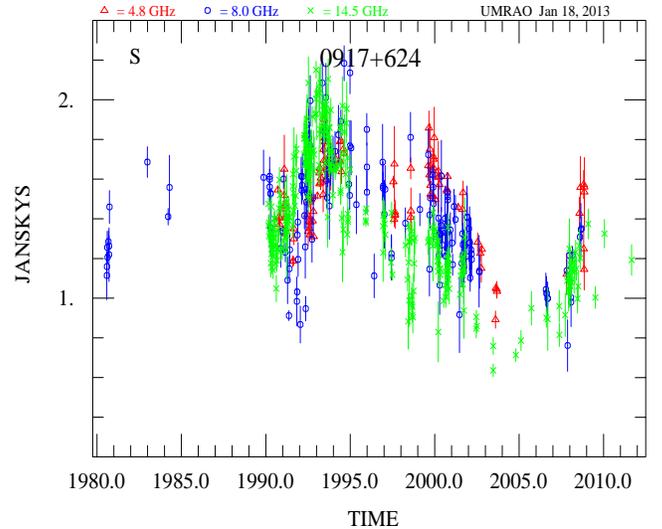}
     \caption{Daily averages of the total flux density versus time at 4.8, 8.0, and 14.5 GHz
     from the University of Michigan Radio Astronomy Observatory (UMRAO). Symbols for the 3 observing frequencies are denoted at the top left.}
     \label{fig1}
 \end{figure}
As an additional test for the presence of IDV, we used the
root-mean-square (rms) total flux density variation normalized by
the mean total flux density as a `variability index' (also called
the modulation index, $m$). We find that the variability index $m$
is $<$ 2\% in all of our observing sessions, with the 1$\sigma$
uncertainty of the total flux density variability
$m_{0}\sim0.5\%$. Here $m_{0}$ is the mean modulation index of all
calibrators (3c48, 3c286, NGC7027, etc). In the majority of the
sessions, the source showed no IDV. However, there appears to be
weak IDV at the 3$\sigma$ level in some of the Urumqi sessions. We
conclude that the strong IDV of 0917+624 which was frequently
detected before 2000 was not
present during our observations. \\

\section{Discussion of the disappearance of the strong IDV}

There are two possible explanations which could account for the
disappearance of the strong ISS-induced IDV. One is that the
interstellar medium which is responsible for the strong IDV of
0917+624 has moved out of our line of sight, or, possibly, that
the properties of the interstellar medium have changed with time.
The second possibility is that the size of the scintillating
component in this core-dominated  source has enlarged and is
greater than the Fresnel angular scale \citep{nar92, wal98,
kri02}, leading to the quenching of the strong IDV.
\begin{centering}
\begin{figure}
     \includegraphics[width=8cm]{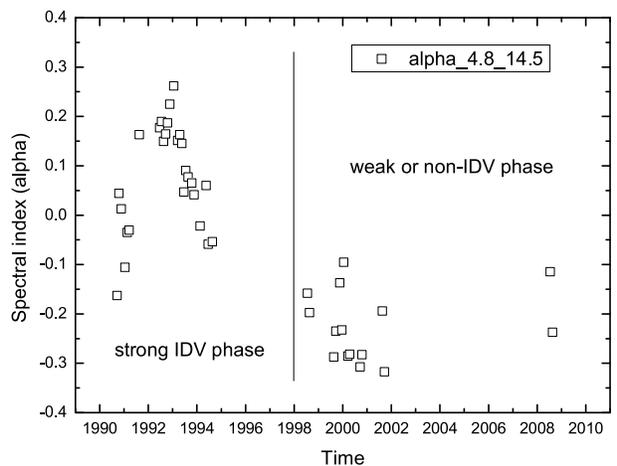}
     \caption{Spectral index between 4.8 and 14.5 GHz from paired monthly-binned UMRAO
     total flux density data versus epoch. $S \propto \nu^{\alpha}$.}
     \label{fig2}
   \end{figure}
\end{centering}
We cannot investigate the first explanation using an in-depth
analysis because we do not know the distance to the interstellar
medium responsible for producing the strong IDV; this was not
well-constrained from the prior ISS model fitting \citep{ric01,
fuh02}, although for a $\sim$1-2 day IDV timescale one obtains
distances of the order of $\sim$100-200 pc. To test the second
scenario, a quenching of the strong IDV due to an enlarged size of
the scintillating component of 0917+624, we
 examined the temporal behavior of
the centimeter-band spectral index using the UMRAO monitoring data
shown in Fig.~\ref{fig1}. In general a steeper spectral index
would imply an enlarged core component, which, probably being
larger than the Fresnel size, could quench the strong IDV of
0917+624. In Fig.~\ref{fig2} we plot the spectral index between
4.8 and 14.5 GHz obtained from paired monthly-averages of the
UMRAO data as a function of time, and we mark the time demarcation
between the weak and strong IDV phases identified in previous
studies with a vertical line. \citet{kra99} found that the strong
IDV of 0917+624 was almost quenched in September 1998 (or that it
has a long timescale of $>$ 5 days), but it was partly recovered
in February 1999; later, the strong IDV ceased again \citep{fuh02,
kra03}. Fig.~\ref{fig2} illustrates a trend of steeper spectral
index after 1998; the strong IDV of 0917+624 occurred mostly
before 1998, and the weak/non IDV phase appeared after 2000. A
correlation between spectral index and ISS was also found for a
large sample of 140 sources by \citet{koa11}. While this behavior
is consistent with our second proposed explanation,
 high resolution VLBI imaging data are needed to confirm whether or not the core was, in fact,
enlarged and whether there is a correlation between the time
of weak/non IDV and the changes of the VLBI core properties.\\

Another possibility is that the most compact ``core'' component
may not have increased in size, but the quasar may have become
less core-dominated, e.g. due to a shock enhancement propagating
down the jet. In that case the scintillation may be no longer
detectable in the single-dish monitoring observations (overall
modulation index $<$2\%), as the flux density is dominated by a
larger (millarcsecond or sub-mas scale) jet component with steeper
spectral index, and possibly the core has faded.

\acknowledgments We thank the anonymous reviewer for valuable
comments. This research has made use of data from the University
of Michigan Radio Astronomy Observatory which has been supported
by the University of Michigan and by a series of grants from the
National Science Foundation, most recently AST-0607523. The work
is supported by the National Natural Science Foundation of China
(Grant No.11073036) and the 973 Program of China (2009CB824800).

\end{document}